\documentclass[letterpaper]{article}

\usepackage[utf8]{inputenc}
\usepackage[margin=3cm]{geometry}
\usepackage{graphicx}
\usepackage{hyperref}
\usepackage{amsmath}
\usepackage{algorithm}
\usepackage{algpseudocode}
\usepackage{float}
\usepackage{array}
\usepackage{booktabs}
\usepackage{wrapfig}

\newtheorem{lemma}{Lemma}
\newtheorem{theorem}{Theorem}

\title{Pcodec: Better Compression for Numerical Sequences}
\author{
Martin Loncaric \\
\texttt{m.w.loncaric@gmail.com}
\and
Niels Jeppesen \\
\texttt{niejep@dtu.dk}
\and
Ben Zinberg \\
\texttt{bzinberg@alum.mit.edu}
}
\date{}

\begin{document}
\maketitle

\begin{abstract}
We present Pcodec (Pco), a format and algorithm for losslessly compressing numerical (float or integer) sequences.
Pco's core and most novel component is a binning algorithm that quickly converges to the true entropy of smoothly, independently, and identically distributed (SIID) integers.
We mathematically prove this convergence with a practical bound.
To accommodate data this is not SIID, Pco has two opinionated preprocessing steps.
The first step, Pco's mode, decomposes the numbers into more smoothly distributed integer latent variables.
The second step, delta encoding, makes the latents more independently and identically distributed.
We demonstrate that Pco achieves 29-94\% higher compression ratio than other numerical codecs on six real-world columnar datasets while using less compression time.
\end{abstract}

\section{Introduction}

Overwhelming quantities of numerical data are stored in columnar, tensor, and time series formats, with even individual datasets measured in petabytes and exabytes \cite{twitter_exabyte,photon_exabyte,tensor_store}.
Compression improvements offer tremendous value in freed data storage.
And since these datasets are often consumed by query engines and machine learning workloads from disk or network devices, better compression can improve performance as well \cite{compression_in_databases}.

Much prior research on lossless numerical compression has focused on speed rather than compression ratio \cite{fast_pfor,fast_lanes,fast_int_simd,general_simd}.
These approaches are particularly suitable for data transfers between CPU cores, where per-core memory bandwidth is high.
However, when reading from disk or network devices, the optimal balance may require stronger compression.
In this regime, CPU can cease to be a bottleneck, leaving compression ratio as the only performance consideration.
Additionally, compression ratio is paramount when compressing for long-term storage, where costs are directly proportional to file size.
Relatively little research has been done to address this need for high compression ratios on general numerical sequences.
Most such approaches we know of use general-purpose LZ77 \cite{lz77} or LZ78 \cite{lz78} (hereafter referred to as \emph{LZ}) codecs as a final step.
We will briefly review several popular or published codecs in this space.

The columnar format Parquet \cite{parquet_encodings} supports a preprocessing step prior to LZ compression.
Examples of Parquet's encodings include \texttt{Dictionary}, which encodes each number as an index into a deduplicated list of unique values, and \texttt{Delta}, which takes differences between nearby numbers (a type of \emph{delta encoding}) and packs them into variable-length integers.
BtrBlocks built on this idea by adding more encodings and using small samples of data to automatically choose a good encoding \cite{btrblocks}.
Uniquely among the codecs listed here, BtrBlocks eshews LZ compression.
The Turbo PFor library offers a variety of preprocessing steps as well, including its namesake Patched Frame of reference algorithm \cite{turbo_pfor}.
PFor algorithms store small numbers using a fixed bit width and store large exceptions separately.
Blosc2, a tensor format, supports the  \texttt{Shuffle}\footnote{Parquet has a similar encoding known as \texttt{ByteStreamSplit}, but does not support it for integer data types.} preprocessing filter, which transposes the data such that all numbers' first bytes come first, followed by their second bytes, etc \cite{blosc}.

Two LZ codecs are worth mentioning.
First, Zstandard (Zstd) \cite{zstd} is an industry-standard LZ codec, used either by itself or paired with one of the aforementioned preprocessing steps.
SPDP is another LZ codec whose parameters were tuned to suit numerical data, and therefore should be used by itself \cite{spdp}.
However, all LZ codecs fundamentally treat bytes as discrete symbols, discarding numerical information.
They have no a priori understanding that the byte 3 is closer to 4 than to 255.
LZ codecs are also unaware of the boundaries between numbers.
This suggests that there may be room for improvement.

We present Pcodec (Pco), a lossless format and algorithm designed to better compress numerical sequences.
The core idea of Pco is to express the input numbers in terms of smoothly (defined in Section \ref{sec:theoretical_results}), independently, and identically distributed (SIID) unsigned integers.
We call these unsigned integers \emph{latents}.
Pco compresses latents with a technique called \emph{binning}, which quickly converges to the true entropy of any SIID data (Section \ref{sec:theoretical_results}).
Contrast this with the core idea of most LZ codecs, which is to express repetitions of the input bytes in terms of independently and identically distributed (IID) symbols from a small alphabet, and to use entropy coding techniques to converge to the Shannon entropy of the symbol distribution.
Intuitively, we expect Pco to outperform LZ codecs on numerical distributions because it is more natural to express numbers in terms of mathematical formulas on latents than in terms of repeated byte sequences.
Pco decomposes numbers into latents via two preprocessing steps: its \emph{mode} and delta encoding.
By default, Pco uses robust algorithms to automatically choose good configurations for these.
This process is opinionated; each chunk of data can use exactly one mode and one delta encoding.

On a wide range of real-world datasets, we find that Pco obtains 29\% to 94\% higher compression ratio than alternatives, even when granting those alternatives 50\% more compression time.
Additionally, Pco's decompression speeds consistently surpass 1GiB/s per thread.

Early work on Pco started in May 2021.
Today, it is used in tools like Zarr \cite{zarr} and CnosDB \cite{cnosdb} to store petabytes of data.
Source code is available at \url{https://github.com/pcodec/pcodec}, including a CLI with benchmarks for all codecs we compare against.

\section{Method}

\begin{figure}
    \centering
    \includegraphics[width=0.75\textwidth]{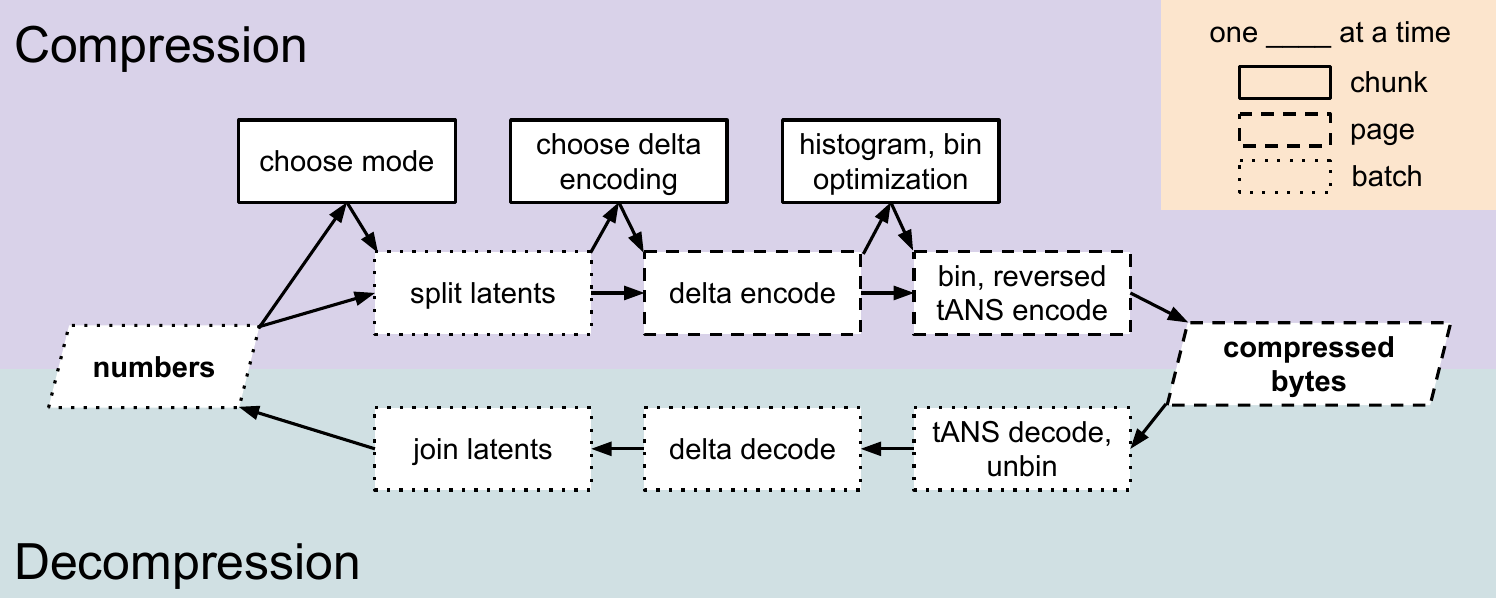}
    \caption{
        The processing steps involved in Pco compression and decompression. 
        By default, compression automatically chooses the mode and delta encoding. 
        Compression must be done one coarse-grained chunk at a time, but can be written out in fine-grained pages. 
        Decompression can be done in even finer-grained batches.
    }
    \label{fig:processing}
\end{figure}

We cover Pco's compression steps in order.
Decompression follows a simpler process with all essential steps done in reverse (Figure \ref{fig:processing}).

\subsection{Modes}

At present, Pco has four modes: \texttt{Classic}, \texttt{IntMult}, \texttt{FloatMult}, and \texttt{FloatQuant} (Table \ref{tab:modes}).
Each mode's purpose is to exploit common structure in the numbers by splitting the distribution into one or two latent variables during compression.
During decompression, the mode joins these latents back into the original numbers.
Although the sum of entropy over latent variables must be at least as high as the original distribution's entropy, the latent variables can be smoother, reducing the number of bits used downstream during binning.

\begin{table}
    \centering
    \begin{tabular}{lll}
        \toprule
        Mode & Types & Example Use Case \\
        \midrule
        \texttt{Classic} & all & normally distributed data \\
        \texttt{IntMult} & integers & ms-precise timestamps stored as \textmu s \\
        \texttt{FloatMult} & floats & prices that are multiples of 0.01 \\
        \texttt{FloatQuant} & floats & 16-bit floats stored as 32-bit floats \\
        \bottomrule
    \end{tabular}
    \caption{Overview of all modes currently supported in Pco.}
    \label{tab:modes}
\end{table}

For brevity, we describe only $\texttt{Classic}$ and $\texttt{IntMult}$ in depth, but the others are similar in principle.
\texttt{Classic} mode is tailored for numbers that are already smoothly distributed.
Therefore, it simply splits each number to a single latent in an order-preserving way.
For instance, a positive float is converted to a latent by toggling its sign bit, while a negative float has all bits toggled (Table \ref{tab:classic_float_examples}).

\begin{table}
    \centering
    \begin{tabular}{ccc}
        \toprule
        Float Value & IEEE Representation & \texttt{Classic} Latent Representation \\
        \midrule
        -NaN & $111\ldots 1$ & $000\ldots 0$ \\
        -2.0 & $110\ldots 0$ & $001\ldots 1$ \\
        -0.0 & $100\ldots 0$ & $011\ldots 1$ \\
        0.0  & $000\ldots 0$ & $100\ldots 0$ \\
        2.0  & $010\ldots 0$ & $110\ldots 0$ \\
        NaN  & $011\ldots 1$ & $111\ldots 1$ \\
        \bottomrule
    \end{tabular}
    \caption{
        Examples of how \texttt{Classic} mode preserves order when converting floats to latents.
        The IEEE representation of each float is non-monotonic in the float's numerical value, but the latent representation is.
    }
    \label{tab:classic_float_examples}
\end{table}

\texttt{IntMult} mode handles integer distributions that resemble a multiply-add of two latent variables, i.e. $x_i = q_im + r_i$ for some parameter $m$ and latent distributions $q_i \sim Q_m$, $r_i \sim R_m$.
For instance, it could be beneficial for a jagged distribution where 90\% of numbers are congruent to 7 modulo 101.
By using $m = 101$, \texttt{IntMult} mode would let $q_i$ be $\lfloor x_i / 101\rfloor$ and $r_i$ be $x_i$'s remainder mod 101.
The resulting $Q_m$ and $R_m$ distributions would be smoother than the original distribution.

\subsubsection{Automatic Detection}

When the user does not configure the mode, Pco chooses it via statistical techniques on a deterministic random sample containing a few percent of the numbers.
Each non-\texttt{Classic} mode chooses a parameter and an estimate of how many bits it would save over $\texttt{Classic}$.
Pco proceeds with the mode with the lowest estimated number of bits required, which we refer to as the \emph{bit cost}, denoted as $\hat{H}$.

Consider \texttt{IntMult} mode.
To propose values for its parameter $m$, we group our sample of $n_\text{sample}$ numbers into $c = \lfloor n_\text{sample} / 3\rfloor$ triples, then let $m = \gcd(x_2 - x_1, x_3 - x_1)$ for each triple $(x_1, x_2, x_3)$.
It follows that $x_2$ is congruent to $x_3$ modulo $m$, so if some value of $m$ appears for anomalously many triples, it suggests that the quotient distribution $Q_m$ might be low-entropy.
With that intuition, we search for frequent values of $m$ among the triples.
In Appendix \ref{sec:derive_int_mult_detection}, we derive conservative approximations for the bit costs of $Q_m$ and $R_m$, leveraging \cite{gcd_probability}.
Let $c_m$ be the number of times the GCD $m$ appeared, and let $n_\text{infrequent}$ be the count of numbers $x_i$ from the sample with at most $n_\text{sample} / 256$ sharing their value for $q_i$.
Then the bit cost of $Q_m$ can be compared to that of \texttt{Classic} mode:
\[\hat{H}[Q_m] - \hat{H}_\texttt{Classic} \approx -n_\text{infrequent} \log_2\left(m\right)\]
And the bit cost of $R_m$ can be measured absolutely:
\[\hat{H}[R_m] \approx \max_{R_m'}H[R_m'], \qquad \text{s.t.} \sum_{r=0}^{m-1} P(R_m' = r)^3 = \min\left(\frac{\zeta(2) c_m}{c}, 1\right)\]
where $\zeta$ is the Riemann zeta function.

We compare this relative bit cost $\hat{H}_{\texttt{IntMult}(m)} - \hat{H}_\texttt{Classic} = \hat{H}[Q_m] - \hat{H}_\texttt{Classic} + \hat{H}[R_m]$ to other values of $m$ to choose the best $m$ parameter, then compare against the bit cost for all other applicable, fully-parametrized modes to choose the best mode.
Both \texttt{FloatMult} and \texttt{FloatQuant} use similarly principled techniques to estimate the bit cost of each latent variable.

\subsection{Delta Encoding}

After applying the mode, Pco uses delta encoding to disentangle correlations between latents within each variable, making them more IID.
At present, Pco has 3 delta encodings: 

\begin{itemize}
\item \texttt{None}: Does nothing.
\item \texttt{Consecutive}: Given an order $o$, $1 \le o \le 7$, takes consecutive differences $o$ times.
\item \texttt{Lookback}: Chooses a lookback $l_i$ for each latent, then transforms each latent $y_i$ to be $y'_i = y_i - y_{i - l_i}$.
The lookbacks are stored as an additional latent variable, and therefore get binned as well.
\texttt{Lookback} allows Pco to capture LZ-like patterns at an approximate number level, as opposed to at an exact byte level.
\end{itemize}

\subsubsection{Automatic Detection}

When the user does not configure delta encoding, Pco chooses it by measuring compressed size of different configurations on a sample of data.
In order to capture correlations between nearby latents, this sample is composed of multiple runs of roughly 100 consecutive numbers each.
Because \texttt{Consecutive} with $o>1$ is rarely advantageous, we try orders $o = 1, \ldots, 7$ sequentially and stop early as soon as the compressed size with $o$ exceeds that of $o - 1$.
Once it has computed the sample's compressed size for each configuration, Pco simply chooses the delta encoding with the lowest size.

\subsection{Binning}

Pco's most essential algorithm is binning, which compresses SIID latent variables.
It is the only processing step in Pco that reduces the byte size of the data.
The overarching idea is to choose histogram bins over the latent variable, then encode each latent by its entropy-coded bin and offset into that bin.

Binning takes as input a vector of latents $(y_1, \ldots, y_n)$.
The first step is to compute a histogram of $k$ tight, exclusive bins $(a_1, b_1, n_1), \allowbreak \ldots, \allowbreak (a_k, b_k, n_k)$, where $a_i$ is a lower bound, $b_i$ is an upper bound, and $n_i$ is the count of latents in the bin.
By tight we mean that for every $i$ there exist some $y_j$ and $y_l$ such that $y_j = a_i, y_l = b_i$, and by exclusive we mean that $b_i < a_{i+1}$.

The second step is to optimize these bins, resulting in $(a'_1, b'_1, n'_1), \allowbreak \ldots, \allowbreak (a'_{k'}, b'_{k'}, n'_{k'})$, which we explain in Section \ref{sec:bin_optimization}.
We build a tANS table \cite{ans} to entropy code the optimized bins, with bin $j$ weighted by $n'_j$.

Finally, we write the $y$ values.
ANS codes must be decoded in the opposite order they are encoded, so we encode in reverse.
For each $y_i$, we find the $j$ such that $a'_j \le y_i \le b'_j$ and compute a tANS code for it.
Then we iterate through the $y$ values forwards in batches of 256.
For each $y_i$ we write out the previously-computed tANS code and record the offset $y_i - a'_j$ as a $\lceil\log_2(b'_j - a'_j + 1)\rceil$-bit unsigned integer.
Once all tANS codes in a batch are written, we write these offsets.

\subsubsection{Bin Optimization}
\label{sec:bin_optimization}

Since each additional bin incurs some metadata storage, it is usually suboptimal to store all histogram bins.
When adjacent bins have similar probability density, they can be combined without increasing the binned data's bit cost by much.
We use a well-known $\mathcal{O}(k^2)$ dynamic programming algorithm \cite{optimal_partitioning} to partition the bins optimally, minimizing bit cost.
This algorithm requires only the bins and a cost function -- in our case, the bit cost $\hat{H}_{i:j}$ associated with an optimized bin formed from unoptimized bins $i$ through $j$:
\[\hat{H}_{i:j} = M + n_{i:j} \cdot (\alpha_{i:j} + \beta_{i:j})\]
Here, $M$ is a constant for the number of bits required to store the bin's metadata,
$n_{i:j} = \sum_{l=i}^j n_l$ is the count of latents in the bin,
$\alpha_{i:j} = \log_2\left(n / n_{i:j}\right)$ is the idealized average entropy coding bits per latent in the bin, and
$\beta_{i:j} = \lceil\log_2(b_j - a_i + 1)\rceil$ is the offset bits per latent in the bin.
Essentially, the bin pays its metadata cost and, for each number in its range, entropy coding and offset cost.
Once the optimal bin partitioning is chosen, we replace each partition's bins $(a_i, b_i, n_i), \ldots, (a_j, b_j, n_j)$ with a single bin
\[(a', b', n') = \left(a_i, b_j, \sum_{l=i}^j n_l\right)\]
This completes bin optimization.

\subsection{Format}

Pco is most flexible within a wrapping format.
We found it important to decouple Pco's goal of good compression from the varied access patterns and logical transformations required by real-world applications, which may include nullability, page statistics, or additional indices.
To support these needs, Pco's API produces three components that can be interleaved as needed.

\begin{figure}
    \centering
    \includegraphics[width=0.52\textwidth]{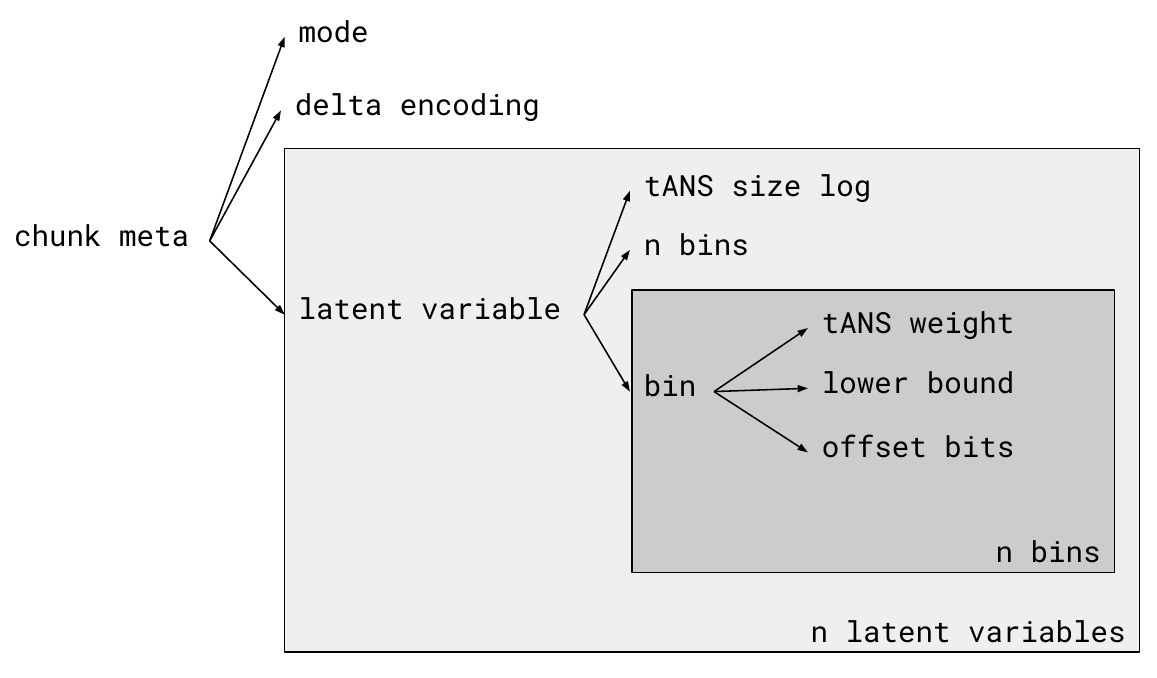}
    \includegraphics[width=0.44\textwidth]{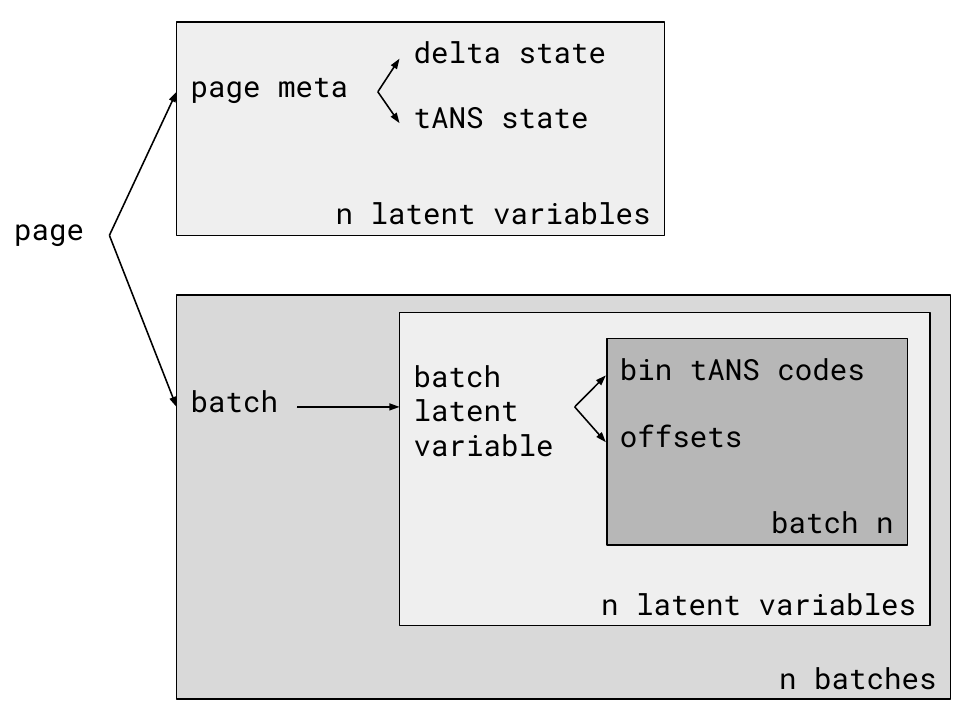}
    \caption{
        Plate notation for the chunk metadata and page components.
        The wrapping format decides where to place each header, chunk metadata, and page.
    }
    \label{fig:format}
\end{figure}

\begin{description}
    \item[Header]
    Pco's header is a single byte and simply indicates Pco's format version, allowing Pco to gracefully add new features in the future.
    Decompressors can always decode older format versions.
    
    \item[Chunk Metadata]
    Each file can contain multiple chunks, the unit of compression.
    In other words, every call to a Pco compression function produces at least one chunk.
    Each chunk has a single mode, delta encoding, and binning configuration, stored in a chunk metadata component (Figure~\ref{fig:format}).
    
    \item[Page]
    Every chunk contains one or more pages, which contain the actual numerical data (Figure \ref{fig:format}).
    Pages store initial states for tANS and delta decoding, so when paired with the correct chunk metadata, they contain all information necessary to decompress their contents.
    Therefore pages can be decompressed in any order; it is not necessary to decompress preceding pages.
    Each page can contain multiple batches, the unit of decompression.
    Batches consist of all latents for 256 numbers and exist solely for performance purposes.
    Batches within a page must be decompressed serially.
\end{description}

\subsection{Performance Considerations}

Speed is Pco's secondary priority after compression ratio, so it is worth mentioning a few techniques that make it fast. First, we compute the histogram via a partial sort rather than full sort.
More precisely, we stop early on quicksort partitions that are entirely within the middle of a bin.
The histogram was originally the slowest part of compression, so this speedup is invaluable, especially when few bins are required.

Next, we vectorize binning by building a binary tree of bins, then iterating primarily by depth into the tree and secondarily by latent in the batch.
During these iterations, we update a bin index for each latent.
This is much faster than the serial algorithm of handling one latent at a time.

Third, we use 4-way interleaved tANS to accelerate entropy coding and decoding.
Entropy decoding is the slowest part of decompression, so this greatly improves decompression speed over the vanilla 1-way algorithm \cite{interleaved_entropy}.

Finally, we vectorize both the bit unpacking and addition of offsets during decompression.
This is possible because we structured each batch of latents as 256 tANS codes followed by 256 offsets, so by the time we reach the offsets, we have full information about which bit range each one occupies.

\section{Results}

\subsection{Theoretical Results}
\label{sec:theoretical_results}

In Appendix A, we obtain a bound on binned bit cost per number:

\begin{theorem}
  Suppose $X$ is a mixture over domain $\{0, \ldots, T - 1\}$ of $s$ disjoint integer distributions, each of which has a monotonic PMF.
  Then for any $k>2s$, there exists a binning of at most $k$ bins such that the expected binned bit cost $\hat{H}$ of a random draw from $X$ satisfies
\[\hat{H} \le H + \frac{3s\log_2(T)}{k-2s}\frac{T}{T-1}\]
where $H$ is the base-2 Shannon entropy of $X$.
\label{thm:bound}
\end{theorem}

Informally, we call $X$ \emph{smooth} if $s$ is small.
And this is the case for many types of data; geometric distributions have $s=1$, Poisson distributions have $s=2$, and float normal distributions have $s=4$ when represented as ordered, unsigned integer latents.
Since the Shannon entropy is the theoretical minimum required average bit cost, this theorem demonstrates $\mathcal{O}(1/k)$ convergence to optimal compression on SIID data.
We prove this theorem by construction, and the binning that attains this bound uses bins of approximately equal probability mass.
This lines up with our approach of using a histogram to assign unoptimized bins.
Note that for any reasonable data type, $T/(T-1) \approx 1$, so in practice the bound is simply
\[\hat{H} \le H + \frac{3s\log_2(T)}{k-2s}.\]

\begin{figure}
    \centering
    \includegraphics[width=0.55\textwidth]{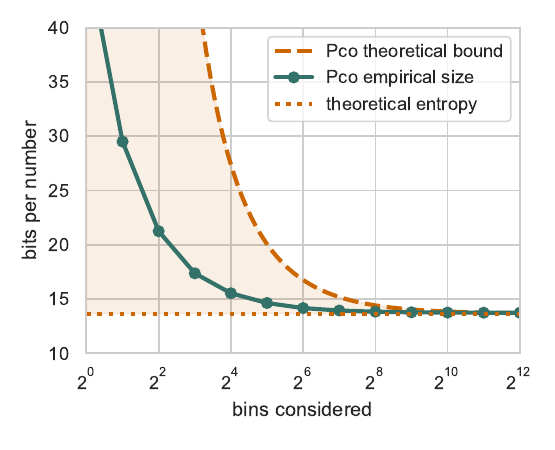}
    \caption{
        Empirical Pco compressed size on a synthetic SIID dataset compared with theoretical bounds: the lower bound of the distribution's true entropy and the upper bound from Theorem~\ref{thm:bound}. 
        One million draws from a Lomax distribution over 64-bit integers were used. 
        Note that our upper bound makes some simplifying approximations and does not account for metadata, but these inaccuracies are very small in practice.
    }
    \label{fig:theoretical_convergence}
\end{figure}

Pco converges faster than this bound in reality (Figure \ref{fig:theoretical_convergence}).
This is to be expected, since our proof uses antagonistic assumptions.
However, this bound can still be practical.
For instance, using the default compression level's 256 bins on a 64-bit data type with $s=1$, we guarantee that binning wastes fewer than $0.76$ bits per latent on average.

\subsection{Experimental Results}

\subsubsection{Setup}

\begin{figure}
    \centering
    \includegraphics[width=1.1\textwidth]{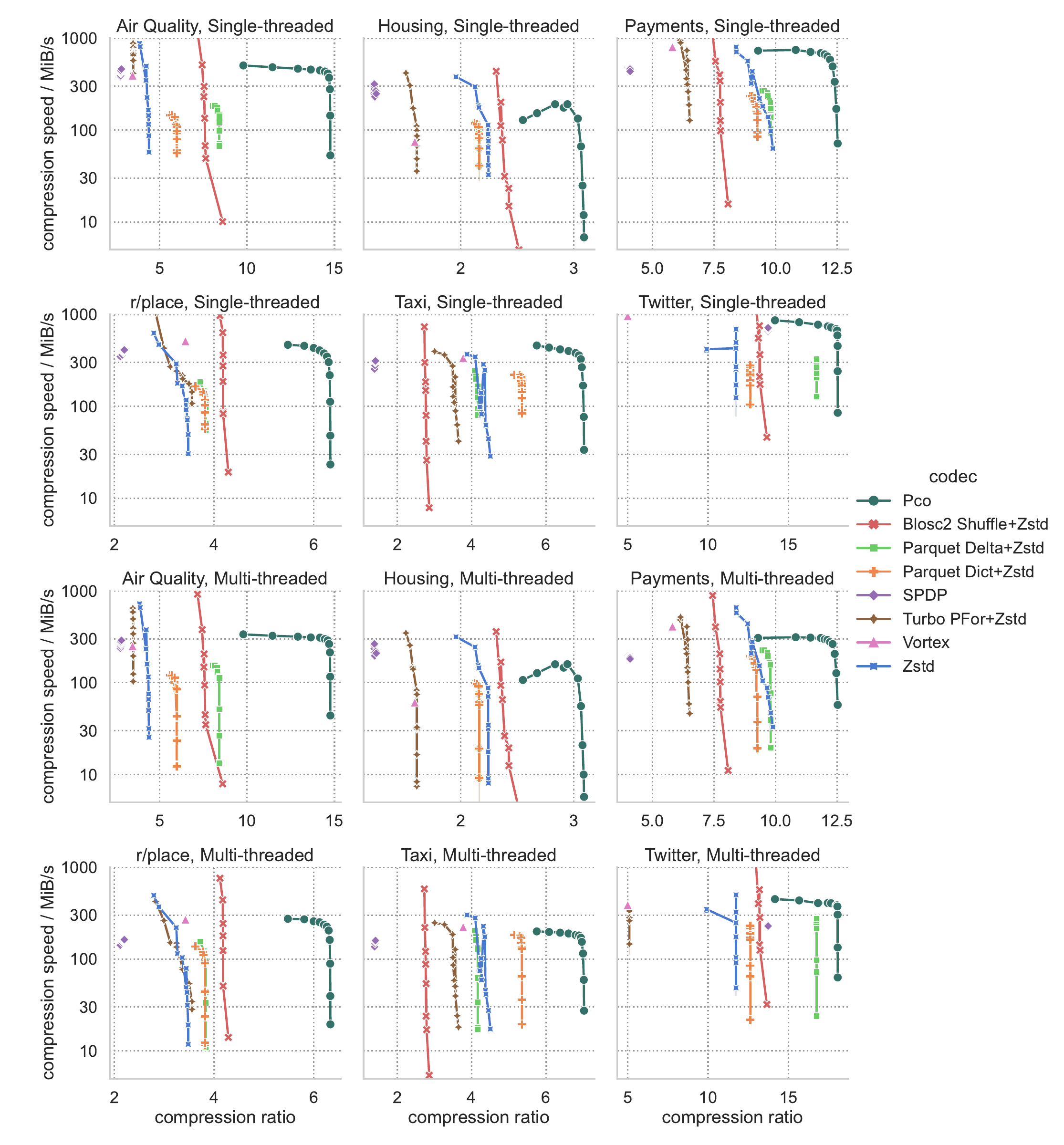}
    \caption{
        Compression characteristics in single- and multi-threaded environments for all codecs on all datasets. 
        In every case, Pco is the Pareto front for its range of compression speeds.
    }
    \label{fig:compression_results}
\end{figure}

We compared Pco versus other popular numerical codecs from industry and literature on integer and float columns from six datasets (Figure \ref{fig:compression_results}).
Each of these datasets are freely available for download:

\begin{description}
    \item[Air Quality] Various air quality metrics from one location, sorted by timestamp \cite{air_quality}.
    
    Data types: \texttt{int32}, \texttt{int64}
    
    \item[Housing] Census data of California housing units \cite{housing}.
    
    Data types: \texttt{float32}
    
    \item[Payments] CMS Open Payments data about medical expenses in the US during 2023 \cite{payments}.
    
    Data types: \texttt{int32}, \texttt{int64}, \texttt{float64}
    
    \item[r/place] Coordinates and colors placed by users during the Reddit Place experiment \cite{r_place}.
    
    Data types: \texttt{int32}, \texttt{int64}
    
    \item[Taxi] NYC Taxi ride data, specifically high volume vehicle-for-hire data from April 2023 \cite{taxi}.
    
    Data types: \texttt{float64}, \texttt{int32}, \texttt{int64}
    
    \item[Twitter] Social graph of twitter users, sorted by IDs \cite{twitter}.
    
    Data types: \texttt{int64}
\end{description}

All benchmarks were performed on a bare-metal 96-core Intel Xeon Platinum 8488C CPU.
In addition to single-threaded benchmarks, we ran multi-threaded benchmarks to simulate moderate load, using 48 cores such that each active core had its own L2 cache.
To minimize measurement noise, we took the median of at least 3 iterations after a warm-up iteration.
We limited large datasets to 2 million rows.

We compared Pco's default configuration versus Blosc2, Parquet, SPDP, Turbo PFor, Vortex \cite{vortex}, and Zstd.
We attempted to compare against BtrBlocks directly, but were unable to on account of segmentation faults.
Instead we added Vortex, an active open source project whose compression strategy is modeled after BtrBlocks.
For Blosc2 we used \texttt{Shuffle}, and for Parquet we considered both \texttt{Delta} and \texttt{Dictionary}.
Since Parquet \texttt{Delta} only applies to integers, we always used \texttt{Dictionary} for floats.
Blosc2, Parquet, and Turbo PFor were paired with Zstd.
We also tested LZ4 and Brotli codecs in place of Zstd, but found that Zstd was strictly better on these datasets.

\begin{table}
    \centering
    \begin{tabular}{cc|rrrrrr}
        \toprule
        Threads & Codec & Air Quality & Taxi & r/place & Housing & Payments & Twitter \\
        \midrule
        1 & Pco & 2,206 & 2,298 & 2,493 & 1,432 & \textbf{3,980} & 5,450 \\
        1 & Blosc2 Shuffle+Zstd & \textbf{4,339} & 2,061 & \textbf{4,435} & 1,416 & 3,800 & \textbf{6,836} \\
        1 & Parquet Delta+Zstd & 2,272 & 2,136 & 2,310 & \textbf{1,679} & 3,728 & 4,910 \\
        1 & Parquet Dict+Zstd & 1,681 & 2,291 & 2,354 & 1,677 & 3,462 & 3,772 \\
        1 & SPDP & 434 & 349 & 391 & 287 & 444 & 539 \\
        1 & Turbo PFor+Zstd & 2,902 & 1,334 & 2,101 & 1,015 & 3,345 & 4,877 \\
        1 & Vortex & 3,455 & \textbf{2,976} & 2,496 & 1,427 & 3,760 & 3,938 \\
        1 & Zstd & 1,550 & 1,248 & 1,221 & 1,069 & 1,973 & 1,350 \\
        \cmidrule{1-8}
        48 & Pco & 1,693 & \textbf{1,566} & 1,696 & 1,186 & 1,818 & 1,961 \\
        48 & Blosc2 Shuffle+Zstd & \textbf{2,930} & 1,311 & \textbf{2,104} & 1,183 & 1,719 & 1,985 \\
        48 & Parquet Delta+Zstd & 1,437 & 1,223 & 1,285 & \textbf{1,394} & \textbf{1,858} & \textbf{2,190} \\
        48 & Parquet Dict+Zstd & 1,057 & 1,287 & 1,270 & 1,391 & 1,712 & 1,792 \\
        48 & SPDP & 254 & 179 & 144 & 241 & 184 & 196 \\
        48 & Turbo PFor+Zstd & 1,301 & 392 & 512 & 833 & 1,055 & 1,344 \\
        48 & Vortex & 1,435 & 1,165 & 576 & 860 & 1,026 & 1,132 \\
        48 & Zstd & 1,178 & 839 & 768 & 893 & 1,132 & 912 \\
        \bottomrule
    \end{tabular}%
    \caption{Single-threaded and multi-threaded decompression speeds per thread in MiB/s.
    Decompression speed varies only slightly between compression levels, so we report its harmonic mean over compression levels.
    Pco consistently decompressed over 1GiB/s, even in multi-threaded settings.}
    \label{tab:decompression_speeds}
\end{table}

\subsubsection{Interpretation}

Given similar compression time, we found that Pco compressed substantially better than all alternatives on all datasets.
Specifically, its default compression level attained 29\% to 94\% higher compression ratio than other codecs, even when allowing them 50\% more compression time (Table \ref{tab:results_improvements}).
In other words, Pco constitutes a 23-48\% reduction in storage costs on these datasets.
These results matched our intuition that binning would suit numerical data better than LZ codecs do.
In the few cases when another codec was able to match the compression ratio of even Pco's lowest compression level considered (2), Pco compressed faster.
In the closest case, Pco at level 2 was 37\% faster than Parquet \texttt{Delta}+Zstd at level 1 on the multi-threaded Payments dataset.

\begin{table}
    \centering
    \begin{tabular}{c|crrr}
        \toprule
        Dataset & Best Alternative & Compression & Compression & Improvement \\
         &  & Ratio (Alternative) & Ratio (Pco) &  \\
        \midrule
        Air Quality & Blosc \texttt{Shuffle}+Zstd (level 4) & 7.54 &14.62 &  +94\% \\
        Housing & Blosc \texttt{Shuffle}+Zstd (level 5) & 2.37 & 3.07 & +29\% \\
        Payments & Zstd (level 4) & 9.06 & 12.21 & +35\% \\
        r/place & Blosc \texttt{Shuffle}+Zstd (level 4) & 4.18 & 6.30 & +51\% \\
        Taxi & Parquet \texttt{Dict}+Zstd (level 2) & 5.29 & 6.94 & +31\% \\
        Twitter & SPDP (level 3) & 13.74 & 18.04 & +31\% \\
        \bottomrule
    \end{tabular}
    \caption{
        The compression ratio of Pco's default level, compared against the next best alternative, allowing for up to 50\% more single-threaded compression time than Pco used.
    }
    \label{tab:results_improvements}
\end{table}

This is not only an improvement in terms of compression ratio, but also user experience.
There is no clear winner among the alternatives we compared, so in absence of Pco, users would need to dynamically choose the best codec for their task at hand.
Pco automatically detects the best compression parameters, reducing complexity.

Pco's decompression speeds were consistently high, exceeding 1GiB/s per thread on all datasets, even in the multi-threaded environment (Table \ref{tab:decompression_speeds}).
This is ample for many workloads that store data or read it from disk or network devices.

\section{Conclusion}

Pcodec is a mature compression format that substantially improves state-of-the-art compression on diverse numerical sequences while operating at practical compression and decompression speeds.
It accomplishes this via a novel binning procedure, distinct from LZ-based approaches.
In addition to leading empirical results, it has strong theoretical results demonstrating rapid convergence to the Shannon entropy of SIID data.

There is room for additional work on two fronts.
First, it would be valuable to quantify the behavior of idealized LZ codecs on SIID numerical sequences.
Comparing this with our theoretical bounds for binning would tell a more complete theoretical story.
Second, although Pco is already used in various wrapping formats, it will take more engineering work to achieve pervasive adoption.

\section{Acknowledgments}

We would like to thank Fabien Giesen for his blog posts and input on accelerating decompression.

\bibliographystyle{ieeetr}
\bibliography{pcodec_arxiv.bib}

\appendix

\section{Derivation of Binning Bit Cost Bound}
\label{sec:theoretical_bound}

We will prove Theorem \ref{thm:bound}, which states:

\

\textbf{Theorem 1} \emph{
   Suppose $X$ is a mixture over domain $\{0, \ldots, T - 1\}$ of $s$ disjoint integer distributions, each of which has a monotonic PMF.
  Then for any $k>2s$, there exists a binning of at most $k$ bins such that the expected binned bit cost $\hat{H}$ of a random draw from $X$ satisfies
\[\hat{H} \le H + \frac{3s\log_2(T)}{k-2s}\frac{T}{T-1}\]
where $H$ is the base-2 Shannon entropy of $X$.
}

\

And less formally, we will show that this binning consists of bins of roughly equal probability.
Also informally, the scaling behavior with imbalanced mixture distributions could be improved in subsection \ref{sec:mixtures}.

To sketch the proof, there will be three main steps.
First, we will show that on any monotonic distribution, there exists a binning in which every nontrivial bin has probability at most $\frac{3}{2(k-1)}$.
Next, we will bound the bit cost when using such a binning.
Finally, we will extend this to mixtures of monotonic distributions.

\subsection{Definitions}

Let $P(x)$ be the PMF of a random integer variable $X$ over a finite, contiguous domain of $T$ distinct $x$ values.
Additionally, let a \emph{binning} with $k$ bins consist of a list of intervals $([a_1, b_1], \ldots, [a_k, b_k])$.
Each such bin has various properties:
\begin{itemize}
  \item \emph{Weight} $w_i = \sum_{x=a_i}^{b_i}P(x)$, the probability mass within the bin.
  \item \emph{Length} $T_i = b_i - a_i + 1$, the number of $x$ values within the bin. A bin is called \emph{trivial} if its length is 1.
  \item \emph{Partial entropy} $H_i = \sum_{x=a_i}^{b_i}-P(x)\log_2(P(x))$.
    It follows that the Shannon entropy $H$ of $X$ is $H = \sum_{i=1}^k H_i$.
\end{itemize}

Finally, define the \emph{bit cost} of a binning to be
\begin{equation}
  \hat{H} = \sum_{i=1}^k w_i(\log_2(T_i) - \log_2(w_i)).
  \label{eq:bit_cost}
\end{equation}

This is a good approximation for the number of bits Pco would use to compress each number in reality.
Because $\hat{H}$ describes an expectation, we can use $\hat{H}$ to model both the bit cost of $X$ and also the expected bit cost of an empirical distribution sampled from $X$.
Note that, for the purposes of theoretical analysis, we have made a few simplifications:
\begin{itemize}
  \item The bit cost of each tANS code is modeled as $-\log_2(w_i)$.
    This is a good approximation, but the true cost may be a fraction of a bit higher on average due to entropy coding imperfections.
  \item The bit cost of each offset is modeled as $\log_2(T_i)$, whereas in reality it is $\lceil\log_2(T_i)\rceil$.
    This difference is not very impactful, especially since our bin optimization step is aware of the true offset cost, and even in antagonistic cases we waste at most one bit per number.
  \item In practice, we must encode the bins themselves as metadata.
    One could add a term of $kM$ onto $\hat{H}$ to account for this metadata cost.
\end{itemize}

\subsection{Constructive Bound on Bin Weight}

\begin{lemma}
  If $X$'s PMF is monotonic, then for any $k$, there exists a binning of $X$ with $k$ or fewer bins such that each bin is either trivial or has weight at most $\frac3{2(k-1)}$.
  \label{lemma:binning_construction}
\end{lemma}

We will prove this constructively.
Suppose without loss of generality that $X$'s PMF is monotonically decreasing.
For any $x$ such that $P(x) > \frac3{2(k-1)}$, dedicate a trivial bin to that single $x$ value.
Suppose there are $k_\text{trivial}$ of these.
Following these, all $k_\text{nontrivial}$ remaining bins must be nontrivial by monotonicity.

Now we use a simple algorithm to assign nontrivial bins greedily, until their weight would exceed $\frac3{2(k-1)}$:
\begin{itemize}
  \item Initialize $w_\text{current} \gets 0$.
  \item For each reamining $x$ value, let $w_\text{proposed} = w_\text{current} + P(x)$, and
    \begin{itemize}
      \item If $w_\text{proposed} \le \frac3{2(k-1)}$, assign $w_\text{current} \gets w_\text{proposed}$ and add $x$ to the current bin.
      \item Otherwise, emit the current bin, start a new one, and set $w_\text{current} \gets P(x)$.
    \end{itemize}
\end{itemize}
Clearly each such bin has weight of at most $\frac3{2(k-1)}$.
And for every bin $[a_i, b_i]$ except the last one, we must have that $w_i + P(b_i + 1) > \frac3{2(k-1)}$; otherwise we would have included $b_i + 1$ in the bin.
Since the bin is nontrivial, it must contain at least two values, implying that $P(b_i - 1) + P(b_i) + P(b_i + 1) > \frac3{2(k-1)}$.
By monotonicity, the first two terms must total at least 2/3 of this quantity, giving $w_i > \frac1{k-1}$.

So we can bound the sum of all trivial and nontrivial bin weights:
\[\sum w_i > \frac{3k_\text{trivial}}{2(k-1)} + \frac{k_\text{nontrivial} - 1}{k-1}\]
\[\sum w_i > \frac{k_\text{trivial} + k_\text{nontrivial}  - 1}{k - 1}\]
The sum of probabilities must be 1, so
\[k - 1 > k_\text{trivial} + k_\text{nontrivial} - 1\]
\[k_\text{trivial} + k_\text{nontrivial} < k\]
In other words, we have used at most $k$ bins, each of which is either trivial or weighs at most $\frac{3}{2(k-1)}$.
This completes the proof of Lemma \ref{lemma:binning_construction}.

\subsection{Bound for Monotonic PMFs}

\begin{lemma}
  If $X$'s PMF is monotonic, then there exists a binning with $k$ or fewer bins such that
  \[\hat{H} - H \le \frac{3\log_2(T)}{k-1}\frac{T}{T-1}.\]
  \label{lemma:monotonic}
\end{lemma}

We will show that the binning from Lemma \ref{lemma:binning_construction} satisfies this.
Without loss of generality, assume $X$'s PMF is monotonically increasing (this is the opposite from the previous section, but makes the indexing simpler).
Consider a bin $[a_i, b_i]$.
The partial entropy of this bin is
\begin{equation}
H_i = \sum_{x=a_i}^{b_i}-P(x)\log_2(P(x)).
\label{eq:partial_entropy_contribution}
\end{equation}
Let $q_i = P(b_i)$, the maximum probability of any $x$ within the bin.
Since $P(x) \le q_i$ within the bin, $-P(x)\log_2(P(x)) \ge -P(x)\log_2\left(q_i\right)$.
Plugging this into Equation \ref{eq:partial_entropy_contribution},
\[H_i \ge \sum_{x=a_i}^{b_i}-P(x)\log_2\left(q_i\right).\]
Which can simply be rewritten as
\[H_i \ge -w_i\log_2(q_i).\]
Supposing that there are $k$ bins, and taking the difference of the total bit cost (\ref{eq:bit_cost}) and these partial entropies,
\[\hat{H} - H \le \sum_{i=1}^k w_i\left(\log_2(T_i) - \log_2(w_i) + \log_2(q_i)\right)\]
By property of the binning we chose and the fact that $P(x)$ is increasing, there is some $k_\text{nontrivial}$ such that bins 1 through $k_\text{nontrivial}$ are nontrivial, and $k_\text{nontrivial} + 1$ through $k$ are trivial.
For the trivial bins, $w_i = q_i$ and $T_i = 1$, leaving
\[\hat{H} - H \le \sum_{i=1}^{k_\text{nontrivial}} w_i\left(\log_2(T_i) - \log_2(w_i) + \log_2(q_i)\right)\]
Now we need only consider nontrivial bins.
We will further divide these into \emph{shallow} and \emph{deep} bins.
Let $k_\text{shallow}$ be the smallest value such that $\sum_{i=1}^{k_\text{shallow}}w_i \ge \frac{3}{2(k-1)(T-1)}$.
We can rewrite the inequality as

\begin{align}
\begin{aligned}
\hat{H} - H \le \sum_{i=1}^{k_\text{shallow}} &w_i\left(\log_2(T_i) - \log_2(w_i) + \log_2(q_i)\right) + \\
\sum_{i=k_\text{shallow} + 1}^{k_\text{nontrivial}} &w_i\left(\log_2(T_i) - \log_2(w_i) + \log_2(q_i)\right)
\label{eq:split_entropies}
\end{aligned}
\end{align}
Since $P(x)$ is increasing, each value of $x$ in bin $i \ge 2$ has probability at least $q_{i - 1}$, so
\[T_i \le \frac{w_i}{q_{i - 1}}, \qquad i \ge 2\]
We can use this to bound the deep portion of Equation \ref{eq:split_entropies},
\begin{equation}
\hat{H} - H \le \sum_{i=1}^{k_\text{shallow}}\left(\hat{H}_i - H_i\right) + \sum_{i=k_\text{shallow} + 1}^{k_\text{nontrivial}} w_i\left(\log_2\left(\frac{w_i}{q_{i - 1}}\right) - \log_2(w_i) + \log_2\left(q_i\right)\right)
\end{equation}
\[\hat{H} - H \le \sum_{i=1}^{k_\text{shallow}}\left(\hat{H}_i - H_i\right) + \sum_{i=k_\text{shallow} + 1}^{k_\text{nontrivial}} w_i\left(\log_2\left(q_i\right) - \log_2\left(q_{i - 1}\right)\right)\]
Since each of these bins has $w_i \le \frac{3}{2(k-1)}$ by Lemma \ref{lemma:binning_construction},
\[\hat{H} - H \le \sum_{i=1}^{k_\text{shallow}}\left(\hat{H}_i - H_i\right) + \frac{3}{2(k-1)}\sum_{i=k_\text{shallow} + 1}^{k_\text{nontrivial}} \log_2\left(q_i\right) - \log_2(q_{i - 1})\]
This sum telescopes, leaving
\[\hat{H} - H \le \sum_{i=1}^{k_\text{shallow}}\left(\hat{H}_i - H_i\right) + \frac{3}{2(k-1)}\left(\log_2\left(q_{k_\text{nontrivial}}\right) - \log_2\left(q_{k_\text{shallow}}\right)\right)\]
Since deep bins are nontrivial, they have $q_i \le \frac{3}{2(k-1)}$, giving
\[\hat{H} - H \le \sum_{i=1}^{k_\text{shallow}}\left(\hat{H}_i - H_i\right) + \frac{3}{2(k-1)}\left(\log_2\left(\frac{3}{2(k-1)}\right) - \log_2\left(q_{k_\text{shallow}}\right)\right)\]

Now we have a reasonable bound for the deep bins, and we can turn our attention to bounding the shallow bins.
All we claim about their partial entropy is that $H_i \le -w_i\log_2(w_i)$, which follows trivially since $w_i \ge P(x)$ for all $x$ in the bin.
This leaves
\[\hat{H} - H \le \sum_{i=1}^{k_\text{shallow}}w_i\log_2(T_i) + \frac{3}{2(k-1)}\left(\log_2\left(\frac{3}{2(k-1)}\right) - \log_2\left(q_{k_\text{shallow}}\right)\right)\]
Let $w_\text{shallow} = \sum_{i=1}^{k_\text{shallow}}$.
Clearly $T_i \le T$, so
\[\hat{H} - H \le w_\text{shallow}\log_2(T) +  \frac{3}{2(k-1)}\left(\log_2\left(\frac{3}{2(k-1)}\right) - \log_2\left(q_{k_\text{shallow}}\right)\right)\]
Also, since mean probability is less than max probability, $q_{k_\text{shallow}} \ge w_\text{shallow}/(T_1 + T_2 + \ldots, T_{k_\text{shallow}}) \ge w_\text{shallow}/T$, giving
\[\hat{H} - H \le w_\text{shallow}\log_2(T) +  \frac{3}{2(k-1)}\left(\log_2\left(\frac{3}{2(k-1)}\right) - \log_2\left(\frac{w_\text{shallow}}{T}\right)\right)\]
\[\hat{H} - H \le w_\text{shallow}\log_2(T) +  \frac{3}{2(k-1)}\left(\log_2\left(\frac{3T}{2(k-1)}\right) - \log_2\left(w_\text{shallow}\right)\right)\]
Thinking of the RHS of this inequality as a function of $w_\text{shallow}$, it is clear there is a single local minimum.
Therefore this bound must be loosest (a maximum on the RHS) when $w_\text{shallow}$ is at an extremal value.
Recall that we selected $k_\text{shallow}$ to be the smallest value such that $\sum_{i=1}^{k_\text{shallow}}w_i \ge \frac{3}{2(k-1)(T-1)}$.
By this choice, and by the limited size of our nontrivial bins, we know that
\[w_\text{shallow} \in \left[\frac{3}{2(k-1)(T-1)}, \frac{3T}{2(k-1)(T-1)}\right]\]
We need only consider these two bounds, $w_s = \frac{1}{T-1}$ and $w_s = \frac{T}{T-1}$.
For convenience, define $w_s$ such that $w_\text{shallow} = \frac{3}{2(k-1)}w_s$.
Our inequality becomes
\[\hat{H} - H \le \frac{3}{2(k-1)}\left((w_s + 1)\log_2(T) - \log_2\left(w_s\right)\right)\]

Consider the $w_s = w_s = \frac{1}{T-1}$ case.
We obtain
\[\hat{H} - H \le \frac{3}{2(k-1)}\left(\frac{T}{T-1}\log_2(T) + \log_2\left(T - 1\right)\right)\]
Similarly, when $w_s = \frac{T}{T-1}$, we get
\[\hat{H} - H \le \frac{3}{2(k-1)}\left(\frac{2T}{T-1}\log_2(T) + \log_2(T-1) - \log_2(T)\right)\]
\[\hat{H} - H \le \frac{3}{2(k-1)}\left(\frac{T}{T-1}\log_2(T) + \log_2(T-1)\right)\]
The result is identical in either case, so our bound holds universally.
We loosen it slightly to make the result more apparent:
\[\hat{H} - H \le \frac{3\log_2(T)}{k-1}\frac{T}{T-1}\]
This completes the proof of Lemma \ref{lemma:monotonic}.

\subsection{Bound for Mixtures}
\label{sec:mixtures}

Suppose $X$ is a mixture of $s$ disjoint, monotonic integer distributions, each with weight $u_j$.
If $k \ge 2s$, we can give mixture component $j$ a total of $k_j = 2 + \lfloor (k-2s)u_j\rfloor$ bins.
Since all $u_j$ sum to 1, the sum of all $k_j$ is at most $k$.
Additionally, $k_j \ge 1 + (k-2s)u_j$.
By Lemma \ref{lemma:monotonic}, this implies a bit cost of at most
\[\hat{H}_j \le H_j + \frac{3\log_2(T)}{(k-2s)u_j}\frac{T}{T-1}\]
The total bit cost is $\hat{H} = \sum_{j=1}^s u_j(\hat{H}_j - \log_2(u_j))$, and the total entropy is $H = \sum_{j=1}^s u_j(H_j - \log_2(u_j))$, so
\[\hat{H} \le H + \sum_{j=1}^s \frac{3\log_2(T)}{(k-2s)}\frac{T}{T-1}\]
\[\hat{H} \le H + \frac{3s\log_2(T)}{k-2s}\frac{T}{T-1}\]
This completes the proof of Theorem 1.

\section{Derivation of \texttt{IntMult} Bit Cost Approximation}
\label{sec:derive_int_mult_detection}

Recall that we defined
\begin{itemize}
  \item $(x_1, x_2, x_3)$: a triple randomly sampled from the input numbers
  \item $m$: $\gcd(x_2 - x_1, x_3 - x_1)$, a proposed value for the parameter to \texttt{IntMult} mode
  \item $c, c_m$: the total count of triples and count of triples with GCD $m$, respectively
  \item $q_i \sim Q_m, r_i \sim R_m$: latents drawn from their respective latent variables, defined as $q_i = \lfloor x_i / m\rfloor, r_i = x_i \mod m$
\end{itemize}

Note that
\[0 \cong x_2 - x_1 \cong x_3 - x_1 \mod m.\]
Adding $x_1$ to all sides gives
\[x_1 \cong x_2 \cong x_3 \mod m.\]
Intuitively, this suggests that abnormally frequent values of $m$ are likely to be good parameters.
We next make estimates of the bit costs of $Q_m$ and $R_m$ for each $m$, which we use to choose the lowest-cost value of $m$.
To that end, we count the number of occurrences $c_m$ of each distinct $m$.

By using some knowledge of binning, we can make an estimate of $\hat{H}[Q_m] - \hat{H}_\texttt{Classic}$, the relative bits saved by binning $Q_m$ as opposed to $X$.
Suppose we use the same bins that \texttt{Classic} would have used on $X$, but mapped into the space of $Q_m$, and that we keep the same bin assignment for each mapped number.
Each bin would still have the same count of numbers, and hence identical entropy coding cost.
Any \texttt{Classic} bin $[a_i, b_i]$ would become $[\lfloor a_i/m\rfloor, \lfloor b_i/m\rfloor]$ in the domain of $Q_m$.
In most cases, this would reduce the number of offset bits by approximately $\log_2(m)$; the only exceptions are the bins that had small domains, with $b_i - a_i < m$.
Call these bins and samples with $b_i - a_i < m$ \emph{frequent}.
Then conservatively assuming we save 0 bits on these nearly-trivial bins, we obtain a reasonable estimate of
\[\hat{H}[Q_m] - \hat{H}_\texttt{Classic} \approx -n_\text{infrequent}'\log_2(m)\]
where $n_\text{infrequent}'$ is the number of samples that would belong to bins with $b_i - a_i \ge m$ in \texttt{Classic} mode.

But histogram computation is computationally expensive, and we can actually avoid choosing bins at by using two more approximations.
First, we replace the condition $b_i - a_i \ge m$ with the condition that $\lfloor a_i / m\rfloor = \lfloor b_i / m\rfloor$.
This is equivalent up to rounding.
This implies that frequent bins contain only a single $x$ value in $Q_m$'s domain.
Our second approximation is that bins will contain only a single $x$ value iff their weight is at least $1/256$.
We use this constant since Pco's default compression level considers up to 256 bins of roughly equal weight.
With these new approximations, we can define $n_\text{infrequent}$ to be the count of samples $x_i$ that share their $q_i$ values with less than $n/256$ samples, and claim that $n_\text{infrequent} \approx n_\text{infrequent}'$.
This leaves the computationally practical approximation that
\[\hat{H}[Q_m] - \hat{H}_\texttt{Classic} \approx -n_\text{infrequent}\log_2(m)\]

For $R_m$, we can actually make a worst-case estimate of the entropy $H[R_m]$ using only $m$, $c$, $c_m$, and the assumption that $Q_m$ is resembles a uniform distribution over a wide domain.
We are constrained by the fact that we found $c_m$ occurrences of $m$, so we assert that
\begin{equation}P(\gcd(x_2 - x_1, x_3 - x_1) = m) \approx c_m / c \label{eq:gcd_freq}\end{equation}
Results from number theory show that in the limit as $n\to\infty$, for $a, b$ drawn uniformly from $1, \ldots, n$,
\[P(0 \cong a \cong b \mod m) = \zeta(2)P(\gcd(a, b) = m)\]
where $\zeta$ is the Riemann zeta function \cite{gcd_probability}.
Since we modeled $Q_m$ as a uniform distribution over a wide domain, we can apply this result to Equation \ref{eq:gcd_freq} to obtain
\[\frac{1}{\zeta(2)}P(0 \cong x_2 - x_1 \cong x_3 - x_1 \mod m) \approx c_m / c\]
\[\frac{1}{\zeta(2)}P(x_1 \cong x_2 \cong x_3 \mod m) \approx c_m / c\]

\[P(x_1 \cong x_2 \cong x_3 \mod m) \approx \min\left(\frac{\zeta(2)c_m}{c}, 1\right)\]
We can evaluate the left-hand side more explicitly by considering all possible values of $r_i = x_i \mod m$:
\[\sum_{r=0}^{m -1}P(R_m' = r)^3 = \min\left(\frac{\zeta(2)c_m}{c}, 1\right)\]
This gives us a very workable constraint.
Now we can simply approximate
\[\hat{H}[R_m] \approx \max_{R_m'}H[R_m'], \qquad \sum_{r=0}^{m-1} P(R_m' = r)^3 = \min\left(\frac{\zeta(2) c_m}{c}, 1\right)\]
This maximum occurs when one value of $r$ has high probability $p$, and the rest are equally likely, giving
\[\hat{H}[R_m] \approx -p\log_2(p) - (m - 1)(1-p)\log_2(1-p),\\
\text{s.t.} \quad p^3 + (m - 1)(1 - p)^3 = \min\left(\frac{\zeta(2) c_m}{c}, 1\right)\]
In practice, we use the method of false position to determine this cubic root $p$ in a quick and numerically stable way.
We then plug it in to obtain $\hat{H}[R_m]$.

\end{document}